\documentclass{aip-cp}

\usepackage[numbers]{natbib}
\usepackage{rotating}
\usepackage{graphicx}

\begin{document}

\title{$XYZ$ States: Theory Overview}

\author[aff1]{Eric S. Swanson\corref{cor1}}

\affil[aff1]{Department of Physics and Astrononmy \\
University of Pittsburgh\\
Pittsburgh PA 15260 USA} 
\corresp[cor1]{Corresponding author: swansone@pitt.edu}

\maketitle

\begin{abstract}
Various ideas associated with exotic hadrons, and hadronic structure in general, are briefly reviewed.
\end{abstract}

\section{INTRODUCTION}

The steadily enlarging collection of unusual heavy hadrons represents a challenge for theory. Will it be possible to find a principle (or a few) that describe the new phenomena? Or will hadronic phenomenology devolve into a sort of astrophysics, where every heavenly object has its own unique interpretation?  This report briefly reviews current ideas in hadronic phenomenology and then concludes with an argument that simple extensions to current models can indeed provide the missing guiding principles that are required to ``understand" the new crop of hadrons.

That this is not an easy process is illustrated by the long history of di-positronium. In 1946, Wheeler suggested that positronium-positronium bound states are possible\cite{wheeler}. That same year Ore proved that the system was unbound, and the next year proved that it was bound\cite{hy}. The {\it observation} of di-positronium had to wait another sixty years\cite{cass}. In this case the process of reducing quantum electrodynamics to an effective nonrelativistic theory is relatively clean -- something that cannot be said of quantum chromodynamics!

The notion that QCD may manifest at low energy with more complicated structure than $q\bar q$ or $qqq$ dates from the beginnings of QCD, with Fritzsch and collaborators pointing out that gluonic degrees of freedom must appear somewhere in the spectrum\cite{gbs}. The possibility of mixed quark and gluonic matter was also realized early\cite{history}, as was the notion of multiquark states\cite{okun}. 

\section{THE LANDSCAPE OF PHENOMENOLOGY}

QCD is notorious for its stubbornness in revealing its secrets. That, and its rather advanced age of 40 years, have led to a proliferation of ideas concerning the emergent low energy behavior of the theory. Some of these are listed in this section.

\subsection{One Gluon Exchange}

The simplest idea concerning quark interactions relies on asymptotic freedom to claim that one gluon exchange should dominate at short interquark distances.  This idea is verified by explicit lattice calculations\cite{jkm}; however, this indicates perturbative behavior at {\it very} small distances of order 0.1 fm or less. At somewhat larger interquark separations the static interaction remains Coulombic but behaves as $\pi/(12 r)$, as is expected from string dynamics\cite{string}. This is our first indication that things are not as simple as might naively appear. Although the notion appears trivial, it can cause trouble elsewhere. For example, the multipole expansion makes use of the quark interaction in a color octet\cite{peskin}, which is assumed to be perturbative. However, it is possible (and likely) that this form of the octet interaction is only valid at very small separations because stringy dynamics is quite robust\cite{ls}.

\subsection{One Gluon Exchange at Higher Order and other Modifications}

One can, of course, attempt to extend the notion of one gluon exchange interactions by computing higher order corrections in perturbation theory. Efforts along these lines date from the early 80s\cite{gup} and expand on contemporary efforts to classify possible quark interactions in terms of gluonic matrix elements\cite{ef}. Such corrections are possibly relevant to phenomenology because  they induce a nontrivial flavor dependence to the interaction:

$$
\delta V(r) = \frac{1}{4r^3} C_F C_A \, \frac{\alpha_s^2}{\pi} \, \log\frac{m_{\bar q}}{m_q}.
$$
Some of the phenomenology of this relevant to the $D_s$ system is worked out in Ref.\cite{ls2}.

Spin-dependence of the interquark interaction is also of interest. This has been pursued for several decades, with phenomenological ideas settling on the notion of ``scalar" confinement coupled to ``vector" gluon exchange\cite{gi}. We note, however, that how spin-dependence manifests itself is far from obvious in QCD, and that it certainly is {\it not} of the form of simple relativistic model interactions. In particular, a scalar confining interaction for baryons implies an {\it anticonfining} interaction for mesons (along with many other problems such as an unstable vacuum).

A better approach is to develop effective field theory capable of addressing this issue. Recent effort in this regard is contained in Ref. \cite{bv}. Alternatively, it is possible to extract spin-dependent interactions in lattice gauge theory\cite{koma}. This leads to strong constraints in the heavy quark sector.

It is possible that other physical mechanisms modify the short range interaction between quarks. For example, pion exchange of the type

$$
V(r) = \frac{g^2}{4\pi} \frac{1}{12 m_q m_{\bar q}} \, \vec \sigma_q \cdot \vec \sigma_{\bar q} \, \vec \lambda^F_q\cdot \vec \lambda^F_{\bar q} \, [ \mu^2 \frac{{\rm e}^{-\mu r}}{r} - 4 \pi \delta(\vec r)]
$$
has been suggested to be important in the baryon spectrum\cite{gloz}. The important phenomenological feature of this interaction is the flavor dependence evidenced in the factor $\lambda^F \cdot \lambda^F$. In a similar fashion, instanton-induced interactions have been incorporated into models of the baryon spectrum\cite{loring}. Unfortunately, our knowledge of the baryon spectrum is too rough to be able to choose between options based on the data alone.

\subsection{Adiabatic Interactions}

There is a long history of lattice gauge computations of static quark interactions. The usefulness of modern calculations and their relevance to the new exotic hadrons is discussed in section 1.3 of Ref. \cite{bric} and Ref. \cite{bb}. Of particular interest is the possibility of determining adiabatic potentials in novel configurations. For example, quark interactions can be measured in the presence of excited gluonic degrees of freedom\cite{jkm}, which is evidently useful for computing the spectrum of heavy quark hybrid mesons. 

Measurements such as this can be used to validate a series of pre-lattice gauge theory ideas concerning gluonic excitations such as the flux tube model\cite{pat}, the constituent vector particle (see the last of Ref. \cite{history}), string excitations (see Giles and Tye in Ref. \cite{history}), and bag models\cite{bag}. For example, recent results strongly indicate that an efficient representations of low lying gluonic degrees of freedom is as a quasiparticle with $J^{PC} = 1^{+-}$\cite{dud}. This single observation essentially eliminates all of the older gluon models.

A minor industry computing adiabatic potentials for a variety of multiquark configurations exists. An early attempt at the $QQ\bar Q \bar Q$ interaction is contained in Ref. \cite{green} and more modern results for four-quark and five-quark configurations exist\cite{5q}. The relatively simple job of determining whether the static three-quark interaction is dominated by two-body interactions or three-body (the ``Mercedes star" interaction) is still not settled. Currently it appears that the interaction is predominantly two-body at short distances and slowly transitions to a three-body interaction as the interquark separation grows\cite{latt-baryon}.

Nascent applications of these ideas to $XYZ$  states are presented in Ref. \cite{bb}.

\subsection{One Pion Exchange}

The possibility that one pion exchange is important at short range has been mentioned already. Here we focus on its evident utility for describing long range interactions. This is, of course, the famous observation of Yukawa, and is underpinned by spontaneous chiral symmetry breaking. Pion (and other meson) exchange forms a cornerstone for longstanding descriptions of nuclear interactions associated with western European cities\cite{nucl}.

Twenty years ago Tornqvist promoted the idea that pion exchange should be taken seriously for describing the interactions between mesons and suggested that a series of hadronic molecules could exist\cite{torn}. This has given rise to a (very) large collection of theoretical speculation concerning possible binding in many meson-meson systems. The most famous of these is the $X(3872)$\cite{X}.

Pion mediated interactions are typically dominated by $P$-wave couplings (for example $N \to N \pi$ and $D^* \to D \pi$). However the possibility of strong $S$-wave couplings also exists. This has been explored by Close and collaborators\cite{cc}.

\subsection{Additional Chiral Dynamics}

As mentioned, pionic interactions are predicated on chiral symmetry breaking and, as pointed out long ago by Weinberg, this means that an effective field theoretic description of low energy pionic interactions is possible. More recently this idea was extended to nuclear interactions (for a review of this large field see Ref. \cite{epel}). An alternative approach has been in the area of ``unitarized chiral  models" in  which model assumptions are used to extend the theory to infinite order in the chiral expansion\cite{ucpt}. While this certainly removes rigor from the effective field theory it opens many phenomena to modelling. For example, a recent application to the LHCb pentaquark is contained in Ref. \cite{roca}.

Chiral symmetry also comes into play in a novel suggestion due to Nowak, Rho, and Zahed\cite{doub}. These authors constructed an effective field theory that combined chiral with heavy quark symmetries. Although not terribly predictive, the model does relate scalar-vector doublets of opposite parity and hence predicts 

$$
m(D_1) - m(D^*) = m(D_0) - m(D).
$$
This prediction works remarkably well for the $D$ and $D_s$ mesons and hence is possibly important in describing the enigmatic $D_s(2317)$\cite{nrz}. Of course other possibilities exist, including novel short flavor dependent interactions mentioned above (Ref. \cite{ls2}) and $DK$ molecular interpretations\cite{barnes}.

In general it is possible that meson exchange contributes to hadronic properties, and that it does so in an important way in some cases. For example, the J\"{u}lich group has argued that triangle diagrams populated by $D$ and $D^*$ mesons dominate  hidden flavor decays of charmonia (such as $\psi(2S) \to \pi J/\psi$)\cite{han}. These diagrams also contribute to the cusp effect, to be discussed below.

\subsection{Novel Quark Degrees of Freedom}

The idea that hadrons can be described by singularities in pionic (or sigma) fields date to work by Skyrme on nuclear interactions\cite{sky}. Several decades later it was suggested that classical pion fields could be created in heavy ion collisions -- a concept that now goes by the name of ``disordered chiral condensate"\cite{dcc}. More relevant to this report is the notion that SU(3) chiral solitons can be ``collectively quantized" and that this is justified in the large $N_c$ limit. Indeed, the large $N_c$ limit permits the use of classical and static hedgehog field configurations and suppresses the effects of quantum fluctuations about these configurations by order $1/N_c$.

These observations were used by Diakonov, Petrov, and Polyakov to predict an exotic decuplet of five-quark baryons\cite{dpp}, the most famous of which was the $\Theta(1530)$. This initiated a storm of activity, initially propelled by low statistics experiments that confirmed the novel prediction. In the end the pentaquark faded away as high statistics experiments were carried out\cite{pq}. Contemporaneously, it was pointed out that the pentaquark excitations are finite in the large $N_c$ limit, which implies that the pentaquark field configuration is not collective, in contrast with the model assumptions\cite{cohen}.

Dubynskiy and Voloshin have recently postulated the concept of ``hadrocharmonium", wherein a compact charmonium state is imbedded in a larger hadron consisting of light degrees of freedom\cite{duby}. The interaction is provided by a color van der Waals force, which is sufficiently strong to provide binding but sufficiently weak to maintain the character of the charmonium component. The mechanism is analogous to older ideas concerning the possible  binding of  charmonia and  nuclei\cite{nucl-cc} and is presumably related to the newly discovered LHCb pentaquark, which was observed in the $p J/\psi$ system\cite{Pc}.
A rich spectrum of hadrocharmonium states is expected,
in particular resonances decaying into $\chi_{cJ}$ and one or two pions, and resonances with charmonium bound to higher excitations of the light hadronic matter.

The idea of ``diquarks" predates QCD\cite{dq1}. The original notion was that a pair of quarks could combine to form a degree of freedom with bosonic quantum numbers in a color (and flavor) $\bar 3$ representation. This idea is often softened these days to refer to a quark-quark correlation in a hadron. In this way a baryon can be thought of as a diquark ($\bar Q$) bound to a quark ($q$) as follows:

$$
(qqq) \to [qq]_{\bar 3} q \to \bar Q q.
$$
Similarly $(q\bar q q \bar q) \to [qq] [\bar q \bar q] \to \bar Q Q$ and 
$(qqqq\bar q) \to [qq] [qq] \bar q \to \bar Q \bar Q \bar q$.

The most famous application of these ideas was to the light scalar nonet, which is well-known for its anomalous properties\cite{jaffe}. More recently the idea has been expanded to all hadrons\cite{selem}, the ``old" pentaquarks\cite{jaffe2}, and the $XYZ$ state\cite{maiani}. The latter effort assumes scalar and vector $[cq]$ diquarks of mass 1933 MeV and combines them to make an infinite spectrum of exotic four-quark states. For example,

\begin{eqnarray}
|0^{++}\rangle &=& |[cq]_S [\bar c \bar q]_S; J=0\rangle \nonumber \\
|0^{++'}\rangle &=& |[cq]_V [\bar c \bar q]_V; J=0\rangle \nonumber \\
|1^{+\pm}\rangle &=& \frac{1}{\sqrt{2}}\Big( |[cq]_S [\bar c \bar q]_V; J=1\rangle \pm |[cq]_V[\bar c\bar q]_S; J=1\rangle\Big).  \nonumber \\
\end{eqnarray}
Evidently, this formalism leads to a large phenomenology that can be liberally applied to the new exotic states.

A recent and novel application in this area is due to Brodsky and Lebed, who propose a new sort of hadronic matter in which reaction dynamics plays an important role\cite{bl}. Consider an electroweak $B$ decay in which a kaon is produced in conjunction with $cd\bar c \bar u$ quarks. These are postulated to coalesce into diquarks $[\bar c \bar u]$ and $[cd]$ that separate rapidly. The diquarks separate to large distance, suppressing the ``fall apart" mode of the putative exotic state. Hadronization occurs at large inter-diquark distance, which enhances to the coupling to excited charmonium, such as $\psi(2S)$, in agreement with experiment for the $Z(4470)$. The model has also been applied to the $P_c$ via the hypothesis $P_c = [\bar c(ud)]_3 [cu]_{\bar 3}$\cite{lebed}.

Notice that the existence of a ``state" relies on the dynamics of the production process. Thus this is a new kind of hadron that hinges crucially on the production and decay of the state. This stands in contrast to the usual 
interpretation of a hadronic state as a stationary eigenstate of the QCD Hamiltonian.

\subsection{Dynamical Confusion}

Many of the $XYZ$ states lie near threshold and therefore are often associated with weakly bound molecular states, in analogue with the $D\bar D^*$ interpretation of the $X(3872)$. Intriguingly, several of the new states lie just {\it above} threshold: $Z_c(3900)$ [$D\bar D^*$], $Z_c(4020)$ [$D^*\bar D^*$], $Z_b(10610)$ [$B\bar B^*$], and $Z_b(10650)$ [$B^*\bar B^*$]. This strongly suggests that these experimental enhancements may be due to threshold cusp effects rather than quark binding.

That something nontrivial can happen at a threshold can be seen in the following two-channel nonrelativistic example. Consider $a\to a$ and $a\to b$ scattering described by the S-matrix

$$
S = \left(\begin{array}{ll}
         \sqrt{1-\rho^2}\, {\rm e}^{2i \delta_a} & i \rho {\rm e}^{i (\delta_a + \delta_b)} \\
          i\rho {\rm e}^{i (\delta_a + \delta_b)} & \sqrt{1-\rho^2}\, {\rm e}^{2 i \delta_b} \end{array} \right).
$$
Near an $S$-wave threshold at $E=E_0$, $\rho^2 \approx 2 c k$ where $c$ is a constant and 
$$
k^2 = 2 \mu_b (E-E_0).
$$
Under these conditions

$$
\sigma( a \to a) \approx \frac{4\pi}{2\mu_a E} \left| \frac{(1-ck)\, {\rm e}^{2 i \delta_a} -1 }{2i}\right|^2 \approx 
\frac{4\pi}{2\mu_a E} (1-ck) \, \sin^2\delta_a.
$$
As the threshold is approached from above the cross section is well-behaved, however $d\sigma/dE \to -\infty$, indicating a discontinuity. Continuing below threshold shows that this can appear as a cusp in the cross section. 

The importance of this simple phenomenon to interpreting hadronic reactions has been stressed for many years by Bugg\cite{bugg}. More recently it has been applied to the $Z_c$ and $Z_b$ states with results that show very good agreement with the experimental data with no need for a resonance assumption\cite{ess}. The model employed in Ref. \cite{ess} was criticized in Ref. \cite{as} as  being acausal; however, it was subsequently shown that nearly identical results can be obtained with a simple unitary and causal model\cite{ess2}. Models similar to those of Ref. \cite{ess} have been applied to the $Z_c$ states but with different modelling assumptions\cite{zc}. Perhaps not surprisingly, these lead to different conclusions about the nature of the $Z_c$ states. Nevertheless, the conclusion appears clear: threshold effects should be accounted for in analyses when their presence is deemed to be important.

\section{CONCLUSIONS}

This brief survey illustrates the breadth of ideas concerning low energy QCD. It is inevitable that many of them will wither away as experimental data are accumulated. Of course, the capabilities of the lattice gauge theory community continue to grow, both in terms of hardware and in algorithmic sophistication. Thus there is hope that definitive answers will eventually be obtained.  This is good news in view of the difficulty in deciphering something as simple as the short range interaction between quarks! 

New $XYZ$ states number in the twenties at present, and undoubtedly more are on the way. It is likely that many of these are statistical fluctuations and will disappear. The remaining demand interpretation. Some physicists think that the $XYZ$ states point to the inadequacy of our current, rather weak, understanding of low energy QCD, and that a radical overhaul of phenomenology must be made.  I am inclined to be both more conservative and more hopeful. It might be that the new states simply highlight the reality of doing experiment high in the spectrum. It is precisely here that gluonic degrees of freedom must make their presence felt. And it is here that the effects of light degrees of freedom must become apparent, especially near thresholds. Perhaps what we are experiencing is merely the opening of quark and gluonic degrees of freedom that QCD demands must happen.

\nocite{*}
\bibliographystyle{aipnum-cp}%

\end{document}